\documentclass[12pt,preprint]{aastex}

\slugcomment{To appear in The Astrophysical Journal}

\shorttitle{Predictions on the power spectrum of clustered EPS}%
\shortauthors{Gonz\'alez-Nuevo et al.}

\begin{document}

\slugcomment{To appear in The Astrophysical Journal}


\title{Predictions on the angular power spectrum\\
of clustered extragalactic point sources at CMB frequencies
\\from flat and all--sky 2D-simulations}


\author{J. Gonz\'alez-Nuevo and L. Toffolatti}
\affil{Departamento de F\'\i{sica}, Universidad de Oviedo, avda. Calvo Sotelo s/n, 33007
Oviedo, Spain}
\email{jgng@pinon.ccu.uniovi.es; toffol@pinon.ccu.uniovi.es\medskip}

\and


\author{F. Arg\"ueso}
\affil{Departamento de Matem\'aticas, Universidad de Oviedo, avda. Calvo Sotelo s/n, 33007 Oviedo, Spain}
\email{argueso@pinon.ccu.uniovi.es}


\vskip -0.8truecm
\begin{abstract}
We present predictions on the angular power spectrum of cosmic
microwave background (CMB) fluctuations due to extragalactic point
sources (EPS) by using a method for simulating realistic 2D
distributions of clustered EPS. Both radio and far--IR selected
source populations are taken into account. To analyze different
clustering scenarios, we exploit angular power spectra of EPS,
$P(k)$, estimated either by data coming from currently available
surveys or by means of theoretical predictions.

By adopting the source number counts predicted by the Toffolatti
et al. (1998) evolution model -- capable of accounting well for
the available data at radio cm wavelengths -- we are able to
reproduce current data on the two--point angular correlation
functions, $w(\theta )$, of radio sources. We can confirm that the
detection of primordial CMB anisotropies is not hampered by
undetected clustered sources at frequencies $\leq 150-200$ GHz. On the other
hand, our current findings show that at higher frequencies the
clustering signal could severely reduce the detectability of
intrinsic CMB anisotropies, thus confirming previous theoretical
predictions. We also show that unsubtracted EPS can account for
the excess signal at high multipoles detected by recent CMB anisotropy
experiments. Moreover, the additional power due to the clustering of
sources gives rise to a small but not negligible contribution to the
same excess signal. As a final result, we also present an example of a
currently feasible {\it realistic map} of EPS at 70 GHz, by taking
into account data on bright detected sources as well as the
previously quoted model for number counts.

Eventually, these simulated sky maps can prove very useful to test
the efficiency of component separation techniques, the capability
of new algorithms for the detection of EPS and the appearance of
non Gaussian signatures in "residue" CMB maps, in the presence of
sources which are not Poisson distributed in the sky.

\end{abstract}





\section{INTRODUCTION}
Astrophysical foregrounds set an unavoidable limitation to precise measurements of primordial cosmic microwave
background temperature fluctuations, ($\Delta T/T$), even at frequencies close to the intensity peak of the
relic radiation. On the other hand, the millimeter region of the e.m. spectrum corresponds to a minimum in the
spectral energy distribution of the Galaxy and of the majority of the extragalactic sources. This minimum occurs
at the cross--over between radio and dust emission, which steeply rises with frequency, and it is only weakly
dependent on the relative intensity of the two components \citep[see, e.g.,][]{dez99,dez03}. Therefore, this
spectral region is the best one for mapping CMB anisotropies.

As for extragalactic point sources -- i.e., galaxies seen as a `point--like' object through the beam, being
their typical projected angular size $\ll \theta_{beam}$ -- their isotropic distribution gives rise to a
contaminating signal which presents the same {\it average} level all over the sky, at a given frequency
\citep{TE96,tof99}. Thus, it can be only reduced by the identification and detection of as many sources as
possible. On the other hand, the relatively large beam sizes and high flux detection limits of current and
forthcoming earth, balloon and space--borne experiments (e.g., BOOMERanG, VSA, CBI, DASI, ACBAR, Archeops, NASA
WMAP and ESA {\it Planck} missions) imply that only few relatively bright sources can be detected and removed
from current as well as future CMB sky maps \citep[see, e.g.,][]{vie03}. Therefore, the contribution of the
highly undetected extragalactic source populations to CMB temperature fluctuations must be accurately estimated
to avoid an unwanted incorrect reconstruction of the angular power spectrum of primordial anisotropies. This
problem is particularly important at intermediate to high multipoles, i.e. $\ell\geq 1000$, where intrinsic CMB
anisotropies are damped.

Many estimates of temperature fluctuations arising from a Poisson distribution of extragalactic sources have
been worked out in the past by \citet{fra89}, \citet{fra91}, \citet{bla93}, \citet{tof95}, \citet{gaw97}. Soon
after, a thorough analysis of the extragalactic foreground contributions to small--scale fluctuations over the
full wavelength range from $\sim 1$ cm to $\sim 300$ $\mu$m, which improved on previous ones, has been presented
by \citet{TO98}. Assuming a Poisson distribution of point sources in the sky, they found that the central
frequency channels of the {\it Planck} mission will be `clean' (i.e, only a few high latitude pixels will be
contaminated by bright undetected sources). As for radio selected extragalactic sources, which contaminate CMB
anisotropies at {\it Planck} Low Frequency Instrument (LFI) channels \citep{man98}, their clustering signal was
found to give a generally small contribution to temperature fluctuations, thanks to the broadness of the local
luminosity function \citep{dun90} and of the redshift distribution of sources which dilute the clustering signal
\citep{TO98, tof99, b&w02}. At higher frequencies, the clustering of far--IR selected dusty galaxies was found
to give a more relevant -- albeit not dominant -- contribution to temperature anisotropies.

On the other hand, recent results coming from the Sub--millimeter Common Use Bolometric Array
\citep[SCUBA,][]{hol99} surveys are giving increasing evidence that many of the sources detected in this
frequency region are ultraluminous star--forming galaxies at $z\geq 2$ \citep[see, e.g.,][]{sco00, dun01,
ivi02}. A detailed evolution model which fully accounts for the SCUBA counts in the framework of hierarchical
clustering scenarios has been recently presented by \citet{gra01, gra04}. In this model, SCUBA sources mainly
correspond to the phases of intense star formation in large spheroidal galaxies at substantial redshift, a
process which should be almost completed at $z\geq 1$. As a consequence, the relevant redshift range is
considerably limited and the dilution of the clustering signal results relatively reduced; correspondingly, the
amplitute of $\Delta T/T$ fluctuations due to clustering is expected to be large. As a preliminary estimate of
this signal, \citet{s&w99} have shown that if the sub--mm (dusty) galaxies detected by SCUBA at 850 $\mu$m do
cluster like Lyman-break galaxies \citep{gia98}, anisotropies due to clustered sources dominate the Poisson ones
at all angular scales for $\nu\geq 250$ GHz. More recently, various papers \citep{mag01, per03, neg04a}
exploited the evolution model of \citet{gra01} to work out precise theoretical predictions on the power spectrum
of CMB temperature anisotropies due to  SCUBA--selected galaxies. In all these works, the authors exploited the
linear and non--linear growth of galaxy clustering with redshift and the evolution of the bias factor
\citep{mat97, mos98}. All these recent studies clearly indicate that dusty galaxies at intermediate-- to
high--redshifts are strongly correlated in the sky and, thus, they are giving rise to a clustering signal which
probably dominates over the Poisson one in almost all {\it Planck} High Frequency Instrument (HFI) channels
\citep{pug98}.

In view of the previously quoted results and, moreover, for having a tool which could help in the analysis of
current as well as future all--sky maps, we decided to try a different approach for estimating the power
spectrum of EPS temperature fluctuations. This approach is based on a fast algorithm capable of simulating
all--sky maps as well as sky patches on which EPS are distributed following a given two--point angular
correlation function, $w(\theta )$. The method, thoroughly explained in Sections 2 and 3 and also discussed,
albeit partially, by \citet{arg03}, works as follows. First of all, a simulated map of Poisson distributed EPS
is created (see, e.g., \citet{TO98}). The only basic requirements are: a) an input model for the number counts;
b) a fixed value for the pixel size. Subsequently, to obtain a map of extragalactic sources whose positions are
correlated in the sky, it is necessary to select the {\it most reliable} angular correlation function for that
particular source population we want to simulate. Then, by working in the Fourier space, we modify the Poisson
density field by introducing the chosen angular power spectrum, $P(k)$, as discussed in Section 2. Finally,
after having applied an inverse Fourier transform to the map, the fluxes of the simulated sources are
distributed in the sky by adopting the same model number counts used in the first step of the process to obtain
the Poisson density field (see Section 3). Eventually, from this map it is easy to estimate the angular power
spectrum of EPS by applying the standard Fourier analysis.

From the very beginning, we focused on 2D sky maps -- thus avoiding the problem of 3D simulations, much more
expensive in terms of CPU time -- for allowing us the creation of a lot of maps, under the most different
assumptions, in relatively short CPU times. Moreover, with the purpose of producing all--sky maps which could be
used by the CMB community in general and by the {\it Planck} Consortia, in particular, we adopted the standard
HEALPIX pixelization scheme of \citet{gor02} throughout the paper. More specifically, we chose a pixel size of
1.718$\times$1.718 arcmin$^2$, i.e. nside=2048 ($\sim 50\times 10^6$ pixels), in the HEALPIX scheme. As for the
flat sky patches, we use the standard angular dimension, $12^{\circ}.8\times 12^{\circ}.8$ deg$^2$, and a pixel
size of 1.5$\times$1.5 arcmin$^2$, unless otherwise stated. In this latter case, our simulated maps are thus
divided in $512\times 512$ pixels.

The outline of the paper is as follows. In \S 2 we briefly discuss the basic formalism focussing on the
assumptions adopted for efficiently simulating maps of clustered EPS. \S 3 will be devoted to describe the model
number counts and the angular correlations functions of extragalactic sources here adopted. In \S 4 we present
our predictions on the power spectra of temperature fluctuations due to the clustering of `known' source
populations. Moreover, we discuss the excess power at high multipole recently detected by the CBI and DASI
experiments. Finally, \S 5 summarizes our main conclusions.

A flat cosmological model with $H_0=70$ km/s/Mpc, $\Omega_{\Lambda}=0.7$ is used throughout the paper. Anyway,
we have to remind that our conclusions are only very weakly dependent on the underlying cosmology given that the
number counts of extragalactic sources are mainly determined by the evolution with cosmic time of the number
density and/or of the emission properties of the underlying source populations \citep[see, e.g.,][]{ dan87,
dun90, TO98}.

\section{BASIC FORMALISM AND SELECTED METHOD}


We remind here that all the predictions discussed in this paper are based on the assumption of `point--like'
sources. As shown by \citet{RF74}, this is a good approximation as far as the angular sizes of sources do not
exceed the beam width. This is generally the case in our simulations, since the adopted pixel size -- i.e., the
resolution element of the map -- is $2.2-3.0$ arcmin$^2$ and there are very few (very bright) low redshift
sources in the microwave sky which show angular sizes greater than $\sim 1$ arcmin. Moreover, the few bright
extended sources will be anyway detected and removed by standard or new detection techniques \citep{TOC98,
vie01, vie03}.

\subsection{2D simulations of clustered point sources: making the density field}

As a preliminary step towards the realization of a flat (a sky patch) or spherical (all--sky) map of clustered
sources, we distribute point sources by adopting a simple Poisson distribution for the number, $n(\bf{x})$, of
sources per pixel. The mean of $n(\bf{x})$ is then $\langle n\rangle=N(>S_{min})/N_{pixels}$, i.e. the average
number of sources per pixel, which is determined by the total number counts $N(>S_{min})$\footnote{As usual,
$N(>S_{min})=\int_{S_{min}}^{S_{lim}}N(S)dS$, where $N(S)$ indicates the differential source counts and
$S_{lim}$ the flux limit for source detection.} at a given frequency $\nu$ (see \S 3). These sky maps have been
already used and discussed in many other papers \citep[e.g.,][]{TO98, dez99, vie01, vie03} and the reader can
refer to them for more details. Therefore, at this first step, the sources are not spatially correlated in the
sky.

We then define the {\it projected} density contrast at a given point (pixel) as
$\delta(\bf{x})={n(\bf{x})-\langle n\rangle\over \langle n\rangle}$, being $\langle n\rangle$ the average number
of sources per pixel. Therefore, the covariance function of the projected density contrast is the usual
two--point angular correlation function \citep[see, e.g.,][]{Peeb93,Pea97,MaSa01}

\begin{equation}
w(\bf{\theta})=\langle\delta(\bf{x})\delta(\bf{x+\theta})\rangle
\end{equation}

where $\bf{\theta}$ is the angular separation in the sky between the two positions and the brackets indicate an
ensemble average. As a following step, we calculate the Fourier transform of the density contrast

\begin{equation}
\delta(\mathbf{k})=\frac{1}{L^2} \int \delta(\mathbf{x}) e^{-i\mathbf{kx}} d\mathbf{x}
\end{equation}

where $L$ is the angular size of the map we are actually using.

It is very easy to show that the angular power  spectrum $\langle |\delta(\bf{k})|^{\rm 2}\rangle$ is the
Fourier transform of the angular correlation function \citep{Pea97}.\footnote{In the case of a standard
power--law, i.e. $w(\theta )=a{\theta }^{-0.8}$, then $P(k)_{cl}=bk^{-1.2}$.} The angular power spectrum depends
only on $k=|\mathbf{k}|$, being the field homogeneous and isotropic.

Thus, the basic idea of our method is to obtain {\it a density field determined by a given power spectrum} or,
equivalently, by the suitable angular correlation function. We do this by calculating the Fourier transform of
the density contrast, $\delta(\bf{k})$, and obtaining its power spectrum, which is constant for all modes,
$P(k)_{Poiss}= const$, if sources are Poisson distributed in the sky. Subsequently, we introduce the chosen
angular power spectrum of correlated sources, $P(k)_{cl}$, suitable for the particular source population adopted
(see \S 3.2.), by applying the following formula

\begin{equation}
\delta_{corr}(\mathbf{k})=\delta(\mathbf{k})\frac{\sqrt{P(k)_{cl}+P(k)_{Poiss}}}
{\sqrt{P(k)_{Poiss}}}
\end{equation}

Then we apply the inverse Fourier transform to $\delta_{corr}(\bf{k})$ which allows us the recovering of a new
density field, $\delta_{new}(\bf{x})$, in which the Poisson term has been modified by $P(k)_{cl}$, i.e. the
Fourier transform of the chosen angular correlation function, $w(\theta )$. Finally, we calculate


\begin{equation}
n_{new}(\bf{x})=\langle {\rm n}\rangle ({\rm 1}+\delta_{new}(\bf{x}))
\end{equation}

which gives the modified number of point sources at each position
in the map, according to the previously calculated new density
field, $\delta_{new}(\bf{x})$. If $P(k)_{cl}=0$ in equation (3) we
obtain, again, a pure Poisson distribution.

Different angular power spectra can be used for the simulations
and they can be derived from the angular correlation $w(\theta )$,
according to the formula of \citet{BE93}

\begin{equation}
P(k)_{cl}=\frac{2\pi}{L^2}\int w(\theta) J_0(k\theta)\,\theta d\theta
\end{equation}

\noindent where $J_0$ is the zeroth-order Bessel function.

In this first approach we have only taken into account the first two moments of the angular distribution, albeit
this assumption cannot give a full description of the statistical properties of the field.\footnote{There are
only few data on higher order moments currently available in the literature and, moreover, this information is
difficult to introduce in the simulations. On the other hand, the agreement found between the predictions of
\citet{arg03} -- obtained by simulations with the same EPSS--2D code -- and the 1-year WMAP data on the reduced
angular bispectrum \citep{kos03} suggests that the excess power introduced by higher order moments has to be
small.}
Anyway, as already pointed out by \citet{arg03}, this procedure proves ``safe'' given that the total number of
bright sources and the number counts remain unchanged after having modified their spatial distribution by the
suitable angular correlation function (see also Section 3.2.1). Moreover, the recovered $w(\theta )$ matches the
input one quite well. This result is displayed in Figure 1: in the upper panel we plot four $P(k)_{cl}$,
calculated by equation (5), chosen as examples to test the procedure (see caption). Among them, we use the well
determined $P(k)_{cl}$ coming from the APM survey \citep{BE93} as the reference one for this test. This angular
power spectrum is better determined than the other ones due to the very great number of galaxies in the survey.
In fact, this $P(k)_{cl}$ is the only one which clearly shows the transition from the linear to the non--linear
regime. In all the other cases, the data are coming from much shallower surveys on a smaller sky area and, thus,
they first converge to a constant value at low multipole. In the lower panel of Figure 1 we show the recovered
$P(k)$, calculated directly from only one all-sky map: the agreement is very good at all scales, with maximum
relative differences $\leq 20$\%.\footnote{At multipoles $\ell\leq 10$ the difference is slightly higher, $\leq
50$\%, as expected, due to the higher sample variance.}
 In this case we plotted the $P(k)$ of a single map to show that the angular power spectrum can be well recovered from each map and it is not the result of
averaging over $n$ simulations. In the same lower panel of Figure 1 it is also plotted the average value and the
$\pm 2\sigma$ confidence interval of the recovered $P(k)$, after 100 simulations (in the case of a flat sky
patch). The result clearly indicate that the method works well: the slope of the power law $P(k)_{cl}\propto
k^{-1.2}$ is recovered with an error $\leq 10$\%.

Notice that the APM $P(k)$ is the most difficult one to reconstruct: in all the other cases the reconstruction
proves easier, since input $P(k)$s are usually fitted by a simple power--law at all scales where they can be
determined.

\begin{figure}
\epsscale{.90} \plotone{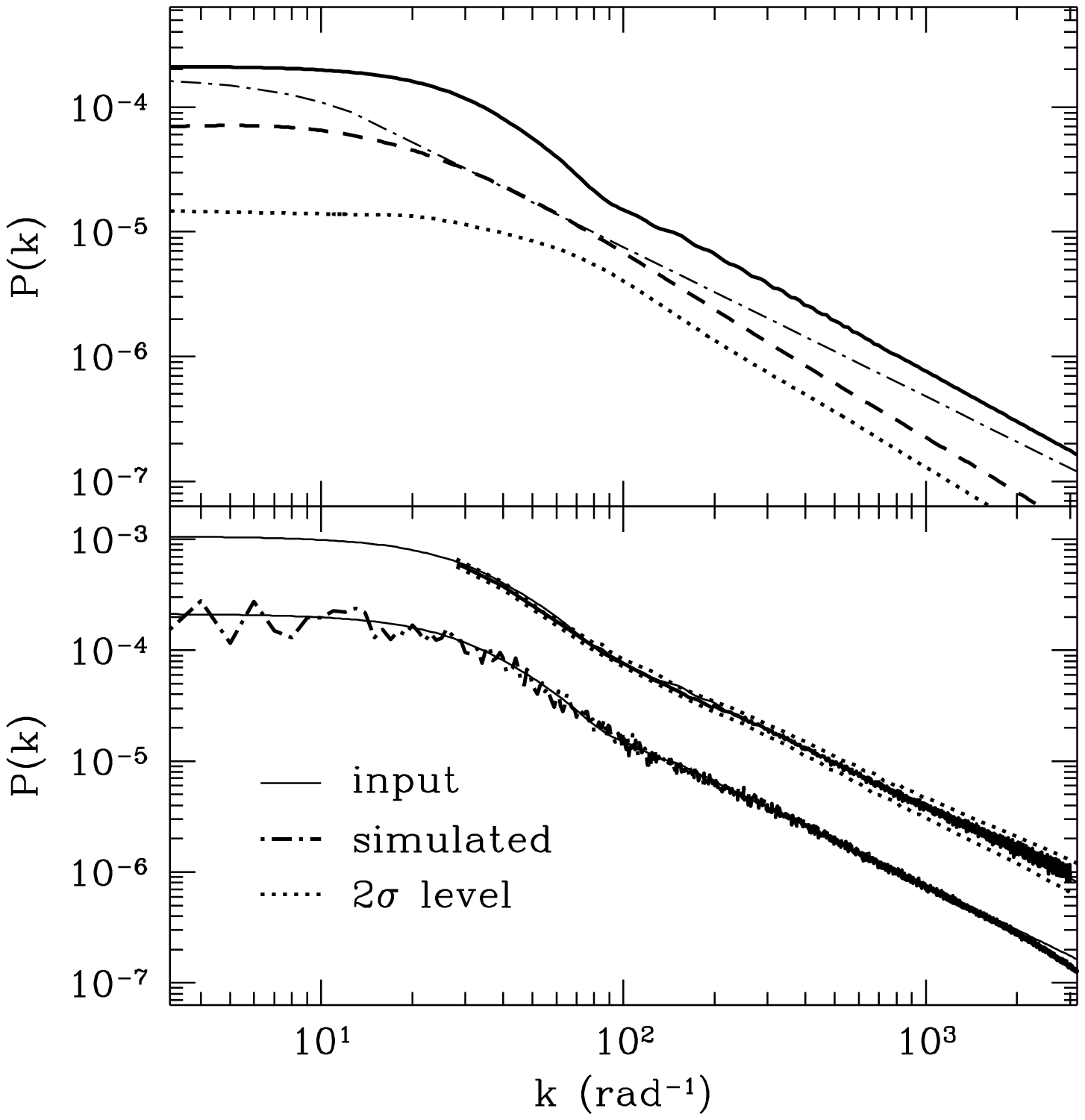} \caption{{\small Upper panel: angular
power spectra calculated by Eq. 5 and from $w(\theta )$ determined
by different surveys. a) the FIRST survey at 1.4 GHz, dotted line,
\citet{mag98}; b) 5 GHz data, dashed--dotted line, \citet{lwl97};
c) optical data of the SDSS survey, dashed line, \citet{teg02}; d)
data from the APM survey, continuous line, \citet{BE93}. Lower
panel: comparison between the input angular power spectrum (thin
continuous line) and the recovered one (dot-dashed line), using
the $P(k)$ calculated from the APM survey data (see \S 2.1.). The
$P(k)$ recovered from a single all-sky simulation is the lower
one. The upper one, arbitrarily shifted for a better
representation, refer to a sky patch. In this latter case we plot
the average value over 100 simulations and the $\pm 2\sigma$
confidence intervals (dotted lines).}}

\end{figure}

\subsection{2D simulations of clustered point sources: how to distribute fluxes}

In the previous subsection we have shown how it is possible to distribute EPS in the sky,  by using their
average number density per pixel, $\langle n\rangle$, which is determined only by the number counts at each
frequency. In this way we have obtained a pure {\it density field}, characterized by the particular $P(k)$ used
in each map and by the average number, $\langle n\rangle$, of sources per pixel. Now, for converting it to a
temperature map, $T(\mathbf{x})$, we have to distribute first the fluxes corresponding to the $N(> S_{min})$
sources of the total counts. Then, the usual conversion $S(Jy) \rightarrow T(K)$ is applied.

Firstly, we have to know the differential counts at a given frequency -- and, in principle, for each source
population -- which are giving us the number, $N(S)$, of extragalactic sources in each flux interval. At each
frequency where it is estimated, $N(S)$ provides a specific distribution function for fluxes, which mainly
depends on the cosmological evolution and on the emission properties of the underlying source populations.
Therefore, it is necessary to find an efficient algorithm for distributing fluxes in the map and which satisfies
two fundamental requirements: a) the differential counts should be recovered within the statistical error
determined by the sample variance; b) the ``input'' angular correlation function has also to be reconstructed,
at least down to the flux limit of the sample by which that specific $P(k)$ has been determined. As already
pointed out by \citet{arg03}, this is a purely {\it phenomenological} approach aimed at simulating all--sky maps
of EPS in a fast way and, at the same time, with the guarantee that the two previously quoted requirements are
satisfied.

In principle, one can choose many different ways of distributing fluxes among pixels: e.g., by simply
distributing fluxes under the condition that in a pixel in which there are $n$ sources we have to put {\it a
corresponding number} $n$ of fluxes, taken {\it at random} from the distribution given by the differential
counts; or, e.g., by imposing that the brightest fluxes fall always in the highest density pixels, thus
strengthening, like imposing a ``bias'' factor, the correlation function. Anyway, no matter the method of
distributing fluxes you choose, the constraint will be always the same: to fulfil the above quoted, a) and b),
fundamental requirements. To achieve this, and with the purpose of simulating maps of EPS at CMB frequencies, we
have first exploited the results coming from the analysis of the NRAO VLA Sky Survey (NVSS) at 1.4 GHz discussed
by \citep{b&w02} (hereafter BW02). By analyzing the great amount of data of this large area survey, the authors
are able not only to quantify the angular distribution of radio sources at this frequency but also to give
estimates of the $w(\theta )$ for different flux ranges. Therefore, by simulating a clustered distribution of
EPS at 1.4 GHz we can test the method more deeply: first of all, for the fact that at this frequency the
differential counts are well constrained down to a few $\mu$Jy and the \citet{TO98}(hereafter TO98) model fits
them quite well down to the faintest fluxes; secondly, for having the possibility of testing the recovered
$w(\theta )$, i.e. the corresponding $P(k)_{cl}$, in different flux ranges.

\subsubsection{Randomly distributed EPS fluxes}

As a first approach, we applied the method of {\it distributing the fluxes at random} among pixels and the
results are shown in Figure 2. In each panel of the Figure we plot the total $C_{\ell}$, given by the Poisson
plus the clustering term; we also represent the input and output $P(k)_{cl}$s (converted to $C_{\ell}$s), due to
the pure clustering term and for different flux limits, following Table 1 of BW02. The power spectra are
calculated from each map by averaging $|\delta(\bf{k})|^2$ over the modes with the same $k$. The $C_{\ell}$s are
then calculated from the $P(k)$s just by multiplying them by the patch area, $L^2$.

Notice that when indicating a flux limit, $S_{lim}$, we do indicate that we have used {\it only those pixels
with resulting fluxes above} the quoted $S_{lim}$ for estimating the $P(k)_{cl}$ from the map. This choice
guarantees that we are correctly comparing the angular correlation function of the simulated map with the input
one of BW02.

\begin{figure}
\epsscale{.90} \plotone{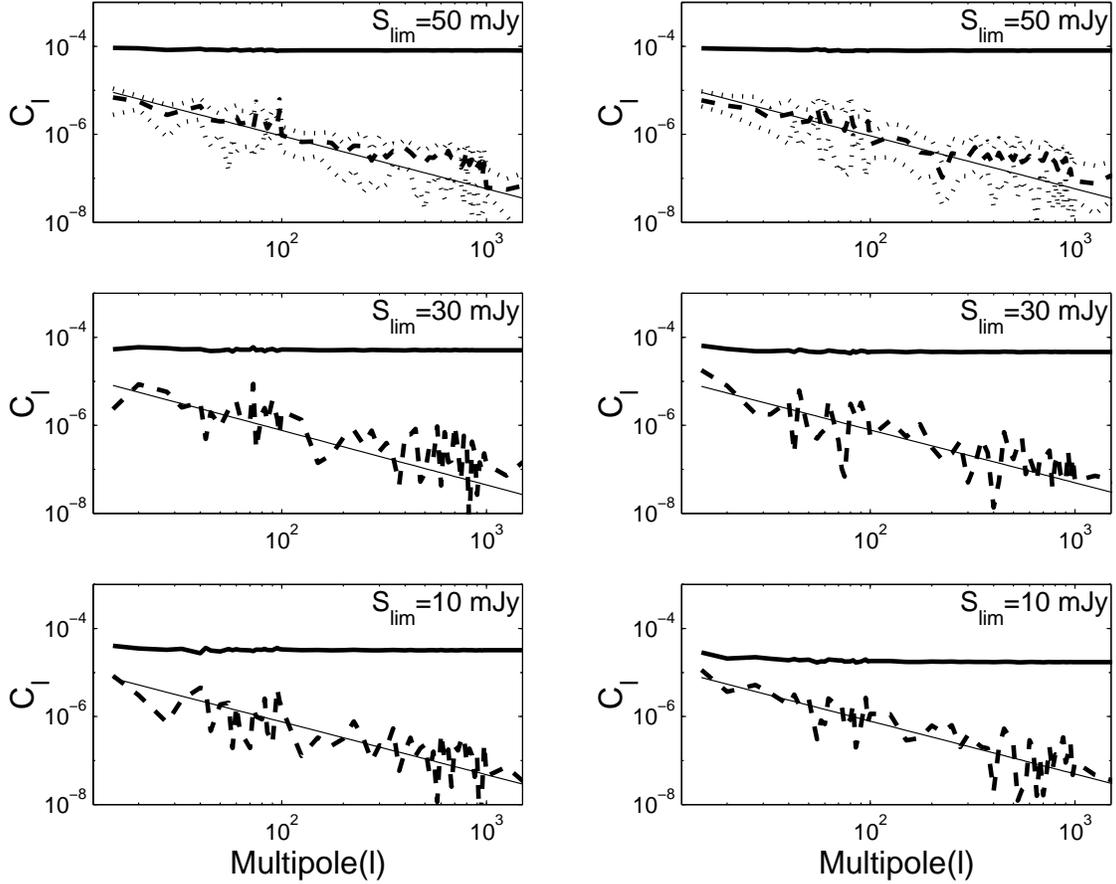} \caption{{\small Left panels: angular
power spectra, $C_{\ell}$ at 1.4 GHz determined for different
source detection limits, $S_{lim}$: 50, 30 and 10 mJy, from top to
bottom. In each panel, the thick continuous line at the top shows
the total power, $P(k)_{Poiss}+P(k)_{cl}$, of the modified power
spectrum; the straight thin line displays the input $P(k)_{cl}$;
and the dashed line, the angular power spectrum, $P(k)_{cl}$,
recovered from the maps. Right panels: in all these cases we have
used only the most accurate value of the amplitude $a=1.04\pm
0.09\times 10^{-3}$ of $w(\theta )=a\theta^{-0.8}$, as given by
BW02. The same $S_{lim}$ values as in the left panels have been
used for recovering the input $P(k)_{cl}$. The plotted results
show the average over 100 simulations of sky patches of
12.8$\times$12.8 deg$^2$. In the case of the two upper panels we
plot the average value after 1000 simulations and the
corresponding $\pm 1\sigma$ confidence intervals.}}
\end{figure}

The maps used for obtaining the results plotted in the {\it left panels} are constructed in the following way: a
single map is obtained by summing up 4 different maps -- with the different amplitudes of $w(\theta )$ given by
BW02 -- corresponding to the flux ranges $S> 50$, $20<S<50$, $1<S<20$ mJy and, finally, a map of Poisson
distributed sources in the range $S_{min}=0.01< S< 1$ mJy, thus reaching the same $S_{min}$ on EPS counts as the
one used by \citet{vie01,vie03}. On the other hand, each final map -- from which the $P(k)$s in the {\it right
panels} of Figure 2. are calculated -- is obtained by only one map in which all sources with $S> 0.01$ mJy are
distributed in the sky according to the most accurate determination of $w(\theta )$ given by BW02 (see caption
to Figure 2). Notice that, {\it in all cases}, we first create the density field, as explained in \S 2.1., and
then distribute {\it at random} the fluxes of the sources previously used for the corresponding density field.
Only sources (and fluxes) corresponding to the specific flux range used, at the chosen frequency (1.4 GHz), are
thus distributed in each map.

From Figure 2 it is clear that: a) the input $P(k)_{cl}$, converted to $C_{\ell}$, is well recovered from the
simulated maps for every chosen $S_{lim}$ in agreement with the original BW02 article, taking into account that
$P(k)_{cl}$ is obtained by subtracting the Poisson term from the total simulated $P(k)$; b) the Poisson term
gives rise to a much greater signal than the clustering term, in all the cases: the total power,
$P(k)_{cl}+P(k)_{Poiss}$, is two/three orders of magnitude above the clustering term. Moreover, the dependence
on $k$ (or $\ell$) of the Poisson and of the clustering term is {\it different}: this well known result is a
direct consequence of the increase of the clustering-to-Poisson ratio at low multipoles \citep{dez96}.
c) the reconstruction is performed with almost the same precision -- average slopes of the $P(k)_{cl}$: $m\simeq
-1.05\pm 0.20$ and $m\simeq -1.13\pm 0.18$, respectively -- by distributing sources using either different
amplitudes, $a$, for different flux ranges (as in Table 1 of BW02) or by using only one amplitude, the one
quoted in the caption of Figure 2. This is not a surprising outcome, given that BW02 state that {\it ``...the
clustering amplitude is also independent of flux''}, like the slope of the power--law, at least in the flux
ranges analyzed by them. Due to this fact, one could question our choice of distributing the faintest sources
(and, thus, fluxes) following a Poisson distribution and not by applying the same power--law used for the
brightest ones. Actually, this choice is not relevant in view of simulating maps for low resolution experiments,
like WMAP and {\it Planck}: the much more numerous fainter sources in the beam (pixel) dilute the clustering
signal and only add some Gaussian noise in the pixel \citep[see, e.g.,][]{bar92,dez96}. On the other hand, it
should be taken into account in the case of simulating maps for the analysis of fields observed with much higher
resolution, e.g. in the optical band.

The results plotted in Figure 2 are quite encouraging and give us confidence in the method. Anyway, for
completeness, we have also checked that if we {\it force} the fluxes of the brightest sources to fall in the
higher density pixels we clearly exceed the input value of the selected $w(\theta )$. This is an expected result
since all our ``input'' angular correlation functions have been determined, as usual, by {\it flux selected}
samples of EPS, and not by samples of objects defined by using some different selection criterion. On the other
hand, if we want to reproduce the distribution in the sky of highly clustered source populations, like the dusty
spheroids at high redshift, discussed in \S 3.2.2, we have to force the brightest sources to fall in the highest
density bins. For each particular source population the correct way of distributing fluxes has to be found, with
the constraint of being able to recover the corresponding observed/theorethical $P(k)_{cl}$ (see \S 4.1).


It is important to stress that the proposed method allows to add up {\it as many source populations as needed}.
However, it shall be generally sufficient to simulate only {\it one source population}, the one which dominate
the counts in the flux interval of interest, except for some very specific frequency channel in which two, or
more, source populations -- showing different clustering properties -- are giving a comparable contribution to
the counts. This is a great advantage, which reduces the total required CPU time, and is determined by the fact
that the total variance of intensity fluctuations due to EPS, $\sigma^2_N$, is obtained by adding up in
quadrature all the contributions coming from different source populations.\footnote{The total {\it r.m.s.
confusion noise}, $\sigma_N$, is given by $\sigma^2_N(S_{lim})=\sum_i (\sigma^2_{P,i}+\sigma^2_{C,i})$, where
the subindexes $i$, $P$ and $C$, indicate a specific source population, and the Poisson and clustering terms,
respectively, and $S_{lim}\leq n\times\sigma_N$ with $n=5$ \citep[see e.g.,][]{neg04a}.} Therefore, the
contribution of faint EPS populations to the total $\sigma_N$ is generally very small.

Eventually, we remind that if we perform the Fourier transform of the temperature map, $T(\mathbf{k})$, the
temperatute power spectrum of the map is calculated by averaging over the modes with the same k values,
$P(k)=\langle |T(\bf{k})|^2\rangle$. Then, in the flat approximation, $C_{\ell}$ can be estimated by
$C_{\ell}=P(k)L^{2}$, where is $L$ is the angular dimension of the sky patch.

As for all--sky maps, the anisotropies of the CMB, ${\delta T/T}$, are usually expanded in spherical harmonics

\begin{equation}
{\delta T\over T}(\vartheta, \varphi)=\sum_{\ell =0}^{\infty}\sum_{m=-\ell}
^{\ell} a_{\ell, m}Y_{\ell}^m(\vartheta, \varphi)
\end{equation}

As usual, the angular power spectrum, $C_{\ell}$, is then calculated averaging the coefficients $|a_{\ell,
m}|^{2}$ with the same $m$. The coefficients $a_{\ell, m}$ are estimated by the routine {\it anafast} of the
HEALPIX package.

We have run our code in a Sun Blade 1000 Workstation UltraSparc III at 750 MHz and with 5 GBytes of RAM memory.
The process of creating one all-sky map -- by distributing only one EPS population -- takes from $\sim$2 to
$\sim$10 hours, depending on the total number of sources ($10^8$-$10^9$), for nside=2048 ($\sim 50\times 10^6$
pixels) in the HEALPIX scheme. The total time is determined by the total number of pixels, $N_{pix}$, and by the
total number of sources, $N_{sour}=N(> S_{min})$ in the map: it goes basically as $O[N_{pix}^{1.5}N_{sour}]$.
The time required to the HEALPIX routine {\it "synfast"} for simulating the pure density field at the same
resolution is $\simeq$ 80 minutes. The rest of the time is determined by the distribution of $\sim 10^8$ up to
$> 10^9$ fluxes, depending on the frequency channel and on the source population. On the other hand, a
512$\times$512 pixels flat patch of the sky is simulated, by a 2.0 GHz processor with 512 Mbytes of RAM memory,
in only a few seconds at 30-100 GHz and in $\simeq 1.5$ minutes at 857 GHz, for one source population.
Obviously, if we distribute in the sky more than one source population, adding up in flux the different
simulated fields, the total time is correspondingly increased.

\section{SOURCE COUNTS AND ANGULAR CORRELATION FUNCTIONS \ AT CMB FREQUENCIES}

\subsection{Number counts of extragalactic sources}

As a benchmark to perform our simulations, we have used the differential counts of the cosmological evolution
model for radio selected sources by TO98 which successfully account for the observed total number counts down to
$S\leq 1$ mJy and up to 8.44 GHz. Moreover, also the relative ratio of ``flat''-to-``steep''-spectrum sources is
well accounted for by the TO98 model (see, e.g., \citet{tuc04}). Recently, new results at 15.2 GHz
\citep{tay01,wal03} suggested that the extrapolation of the TO98 model counts overestimates their data by a
factor of $\sim 1.5\div 1.8$ at the survey limit ($\simeq 20$ mJy). On the other hand, the recently published
analysis of the foreground emission in the Wilkinson Microwave Anisotropy Probe (WMAP) first year data
\citep{ben03b} has provided a full sky catalogue of 208 bright extragalactic sources with fluxes $S\geq 0.9-1.0$
Jy, of which only five objects could be spurious identifications. The whole sample provides an average ``flat''
($\alpha=0.0\pm 0.2$) energy spectrum in full agreement with the assumptions of TO98, and number counts of
bright sources at 41 GHz which appear to fall below the prediction of TO98 at 44 GHz by a factor $\sim 0.7$.
However, a direct calculation of the 33 GHz counts by the source catalogue in Table 5 of \citet{ben03b} gives
155 sources at $S_{lim}\simeq 1.25$ Jy on a sky area of 10.38 sr ($\vert b^{II}\vert> 10^{\circ}$) where the
sample appears to be statistically complete.\footnote{Only four sources are detected at galactic latitude $\vert
b^{II}\vert< 10^{\circ}$ according to Table 5 of \citet{ben03b}. Moreover, by a re-binning of the data in the
same Table we found that the differential counts of WMAP detected sources sink down below $S\simeq 1.25$ Jy,
suggesting the onset of incompleteness.} With this latter source selection the average offset, $\sim 0.75$, with
the TO98 model predictions is slightly reduced.

Furthermore, two other recently published independent samples of extragalactic sources at 31 and 34 GHz -- from
CBI \citep{mas03} and VSA \citep{tay03} experiments, respectively -- show that the TO98 model {\it correctly}
predicts number counts down to, at least, $S\simeq 10$ mJy (cf. Fig. 13 of \citet{ben03b}). As pointed out by
\citet{her04}, it is also noticeable that the brightest extragalactic source in the WMAP first year catalogue
has a flux of $S\simeq 25$ Jy at 41 GHz which, again, corresponds to the flux for which the TO98 model predicts
1 source all over the sky. Another important result is that a good agreement is currently found between
predictions based on the TO98 model and WMAP data on the excess angular power spectrum at small angular scales
\citep{hin03} as well as on the angular bispectrum detected in the WMAP Q and V bands \citep{kos03,arg03}.

On the other hand, a lot of new data on far--IR/sub--mm source counts have piled up since 1998, in particular by
means of the SCUBA and MAMBO surveys. These current data are better explained by new physical evolution models
of far--IR selected sources \citep{gra01,gra04}. These models foresee a steeper slope of the differential number
counts of EPS than the TO98 model do at intermediate fluxes, i.e. $S\simeq 10-100$ mJy, where the contribution
of high redshift spheroids show up \citep[see, e.g.,][]{per03}. In view of this, and for comparing our current
estimates on the $C_{\ell}$ of EPS with the theorethical predictions of \citet{per03}, we have also used the
model counts of \citet{gra01} at the frequencies of 217 and 353 GHz. These are relevant frequencies, which
correspond to the central channels of {\it Planck} HFI, and where the contribution of dusty galaxies to CMB
anisotropies should start to dominate over the one coming from radio selected sources.

\subsection{Angular correlation functions at CMB frequencies}

As for the ``input'' angular correlation function, we have to use, obviously, the $w(\theta )$ most appropriate
to that specific source population we have to simulate (see also \S 2.2.). In this paper we are mostly
interested in simulating all--sky maps at CMB frequencies in view of the optimization of current simulations and
future data analysis for the {\it Planck} mission. In particular we would like to develop a useful tool for
testing the efficiency of algorithms for source detection, for component separation, for detecting non
Gaussianity in the residual maps and, finally, for better denoising of current as well as of future maps. Given
the lack of determinations at CMB frequencies, we have, again, to rely on $w(\theta )$s determined at
frequencies close to the ones under study.

Whereas in the optical domain the very deep large--area surveys of the last decade have provided a lot of very
precise determinations of the $w(\theta )$ -- e.g., APM, 2dFGRS and SDSS sky surveys \citep{BE93,col01,teg02} --
only a few determinations of the $w(\theta )$ at radio (cm) frequencies are currently available
\citep{lwl97,mag98,b&w02,mag04}. Moreover, at HFI frequencies, the only data up--to--now available on clustering
properties of EPS are coming from the SCUBA sky fields. These fields are of very small angular dimensions and,
thus, no clear determination of the $w(\theta )$ at these frequencies has been yet published. Some theoretical
analysis of the clustering properties of SCUBA galaxies have been recently attempted
\citep{knox01,per03,neg04a}.

For all the above reasons, and taking into account previous analyses on number counts of
EPS at CMB frequencies \citep{TO98,gui98,sok01,par02,gra01,per03}, we have made the following choices as for the most reliable $w(\theta )$s to be applied in simulations at
{\it Planck} LFI and HFI frequencies.

\subsubsection{Radio selected extragalactic sources}

All the most recent surveys of radio sources are confirming -- with a small offset -- the predictions on number
counts made by TO98, at least up to $\sim 40$ GHz (\S 3.1.). At these frequencies, the EPS relevant at fluxes of
interest for current as well as future CMB anisotropy experiments are mainly ``flat''---spectrum sources, i.e.
QSOs, BL Lacs and local AGNs, and, thus, the contributions coming from other fainter source populations can be
neglected. Moreover, the number of ``inverted'' spectrum sources -- which could have represented a threat for
CMB $\Delta T/T$ measurements -- is found to be always small \citep{dez00}. Therefore, we could be confident in
adopting the angular correlation function determined by the Parkes-MIT-NRAO survey at 5 GHz \citep{lwl97} for
simulating sky maps of clustered EPS sources at {\it Planck} LFI frequencies and angular resolutions. On the
other hand, when analyzing CMB sky maps coming from high resolution experiments -- which probe fainter fluxes --
one can rely on the $w(\theta )$ estimated by \citet{b&w02} from the NVSS survey. This $w(\theta )$ refer to
fainter sources than the \citet{lwl97} one and represents a more realistic approximation to the clustering
properties of CBI and BIMA radio sources (see \S 4.2).


\begin{figure}
\plotone{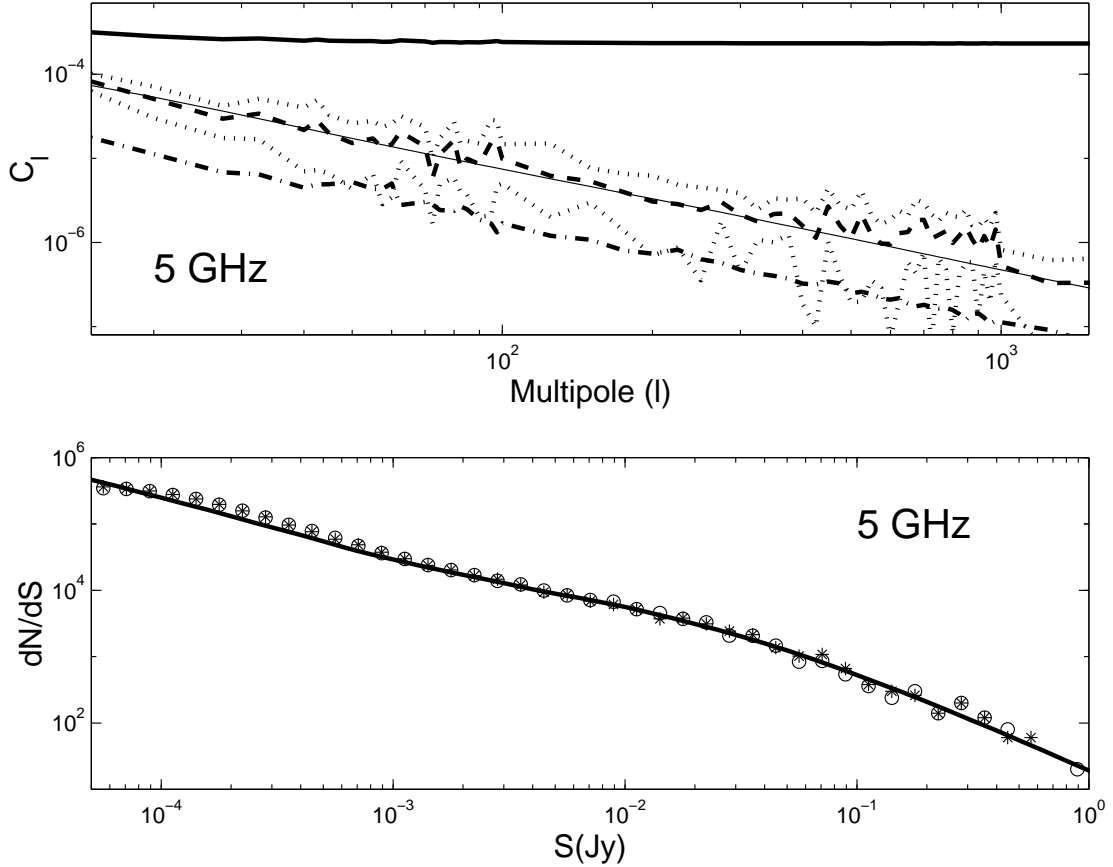} \caption{{\small Top panel: input $P(k)_{cl}$ of \citet{lwl97}, thin continuous line, and recovered
one, dashed line, by using a detection limit $S_{lim}=50$ mJy. The $\pm 1\sigma$ confidence intervals over 100
simulations are also plotted (dotted lines). The thick continuous line at the top of the panel shows the total
power, $P(k)_{Poiss}+P(k)_{cl}$, of the modified power spectrum. The other curve, which appears at a lower level
than the input one, show the recovered $P(k)_{cl}$, having lowered the detection limit down to 0.1 mJy. Bottom
panel: differential counts per $sr$ and per logarithmic flux interval -- $dN/dS=N(S)(\Delta log_{10}S)$, where
$\Delta=0.1$ -- predicted by the TO98 model at 5 GHz (thin continuous line) compared with counts recovered from
one single map: clustered (asterisks) and Poisson distributed EPS (empty circles). Notice that the simulation,
in the Poisson and non-Poisson case, refer to only one sky patch; numbers have been converted to 1 sr.}}

\end{figure}

In Figure 3, bottom panel, we plot the input differential counts of TO98 model at 5 GHz, which match very well
the observed ones (as shown by Figure 1 of TO98), and they are compared to the ones recovered from only {\it
one} simulated map.\footnote{As in the previous and in the next \S, we always distribute all the sources and,
thus, their corresponding fluxes -- foreseen by that specific $N(S)$ distribution  -- down to $S_{min}=0.01$
mJy.} As it is apparent from Figure 3, the recovering of the differential counts from the simulated maps is
achieved with very high precision down to very faint fluxes. In both cases, i.e. Poisson and non-Poisson, the
counts recovered form the map start to exceed the input ones by a small factor only at fluxes $S\sim 0.4-0.5$
mJy, whereas they fall rapidly below the input model counts at $S\simeq 0.05-0.06$ mJy.\footnote{The differences
seen at bright fluxes are easily explained by statistical fluctuations around the mean value in each flux bin,
due to the low number of EPS in these bins. The maximum difference, at 300 mJy, corresponds to the $1.2\sigma$
level (10 sources to be compared with 6, in a flat sky patch).}

Source counts are increased at flux levels much fainter than the flux detection threshold,
$S_{lim}=5\sigma_N\simeq 1.5$ mJy, calculated at 5 GHz by using the same formalism as in \citet{fra89}, TO98 for
a top--hat square beam of FWHM$\simeq 1.5$ arcmin and without any other noise added. Therefore, the adopted
method does not modify number counts -- at least down to the flux level at which EPS can be detected with our
resolution element -- and the clustering term, i.e. the adopted input $w(\theta )$, only negligibly increases
the counts at faint fluxes, as expected in this frequency range.\footnote{At higher frequencies, in the presence
of strongly clustered source populations like high-redshift spheroids, and with low-resolution experiments, the
number counts can be greatly modified by the clustering term, even at the $5\sigma$ level \citep{neg04b}.}

In the case of Poisson distributed sources (empty circles) the simulated counts fall below the input ones at
$S\leq 0.04$ mJy, the level at which the TO98 model predicts $\sim$1 source/pixel, with a pixel size $1.5\times
1.5$ arcmin$^2$. In the case of clustered sources, with the $P(k)_{cl}$ here adopted, the limit of 1
source/pixel is reached at slightly brighter fluxes, i.e. $S\simeq 0.06$ mJy.\footnote{As a final check, we have
also verified that the total flux is conserved: the excess flux measured at $0.07<S<0.5$ mJy is exactly the same
as the one lost at fainter fluxes, due to the effect of source confusion with the fast convergence of the counts
at $S\leq 0.06$ mJy.}


In the top panel of Figure 3 we plot the input and recovered $P(k)_{cl}$ of \citet{lwl97}. At 5 GHz, given the
shallowness of the survey, we can rely only on a preliminary estimate of $w(\theta )\simeq 0.01\theta^{-0.8}$
for fluxes $S\geq 50$ mJy and we can only apply our method by using a single $w(\theta )$. Therefore, we have
distributed EPS in the sky in the same way as at 1.4 GHz but as in the case whose results are plotted in the
{\it right panels} of Figure 2 (only one $w(\theta )$ for all sources at $S> 0.01$ mJy). As for the angular
correlation function, the curves plotted in the top panel of Figure 3. show, again, that the EPSS--2D algorithm
is capable of simulating maps of clustered EPS in which the input $P(k)_{cl}$ is well recovered. We chose to
represent the result of 1000 simulations (and the corresponding $\pm 1\sigma$ intervals), given that the reduced
number of pixels with $S\geq 50$ mJy implies that by only one realization of the sky the simulated $P(k)$
results too noisy (the Poisson term always dominates, as clearly shown by the upper thick line). Anyway, from
100 simulations only we recover an average slope of $P(k)_{cl}$, $m\simeq -1.15\pm 0.1$ and an amplitude whose
offset with the input one is $\simeq 20\div 30$\% and always compatible at the $1\sigma$ level. The other
plotted curve (see caption to Figure 3) shows the effect of the dilution of the clustering signal, when lowering
down the flux detection limit, as already discussed in \S 2.2.


\subsubsection{Far--IR selected extragalactic sources}

For simulating sky maps of far--infrared (far--IR) selected EPS, which dominate the total number counts at
$\nu\geq 300$ GHz, we have relied on the new physical evolution model for sources recently proposed by
\citet{gra01,gra04}. This model, which successfully reproduces the observed counts and the available information
on the redshift distribution of SCUBA sources, describes SCUBA sources as intermediate-- and high--redshift
proto-spheroidal galaxies mainly observed during their principal burst of star formation, whose evolution and
duration is substantially affected by the growth and by the activity of a central super--massive black hole and
not only by supernova feedback. Obviously, spheroidal galaxies (spheroids) are not the only source population
contributing to number counts at far--IR and sub--millimeter wavebands. Late--type galaxies, i.e. normal spiral,
irregular and starburst galaxies, dominate the bright tail of number counts whereas spheroids provide, by far,
the overwhelming contribution to $N(S)$ in {\it Planck} HFI channels for $S\leq 100$ mJy . On the other hand,
late--type galaxies are relatively weakly clustered \citep[see, e.g.,][]{mad03} and, thus, their contribution to
confusion noise turns out to be dominated by Poisson fluctuations \citep{neg04a}.

SCUBA sources are expected to show a strong spatial clustering, similar to the one measured for Extremely Red
Objects \citep[EROs;][]{dad01,dad03}. On the other hand, direct measurements of clustering properties of these
sources is made difficult by the poor statistics. However, \citet{pea00} found some evidence for clustering of
the background source population in the Hubble Deep Field observed at 850 $\mu$m and tentative evidences of
strong clustering, consistent with that found for EROs, have been recently reported \citep{web03}. The power in
excess over Poisson fluctuations detected by \citet{pea00} is well accounted for by a two--point angular
correlation function of the form $w(\theta )=(\theta/\theta_0 )^{-0.8}$ with $\theta_0$ in the range 1--2
arcsec.

As discussed by \citet{per03,neg04a}, the above value of $\theta_0$ is consistent with a number of data on
clustering of SCUBA sources and can be well accounted for by a $w(\theta )$ derived from physical assumptions on
the evolution of clustering \citep{mat97}. In the framework of this physical evolution model for clustering,
\citet{per03} worked out a complete set of predictions on the angular power spectrum due to EPS sources at HFI
frequencies. Given that we are currently interested in testing our algorithm for distributing clustered sources
in CMB sky maps, in the following we adopt the $w(\theta )$ of \citet{per03} and the number counts of spheroids
of \citet{gra01} for simulating all-sky maps of far--IR selected EPS. Therefore, from these maps we have to
recover the same theoretical $C_{\ell}$ values of \citet{per03}.


\section{PREDICTIONS ON THE ANGULAR POWER SPECTRA OF CLUSTERED EXTRAGALACTIC SOURCES}

\subsection{Temperature power spectrum estimates from simulated maps at microwave frequencies}

Figure 4 and 5 display our current results on the temperature angular power spectrum, $C_{\ell}$, due to
clustered EPS at the \emph{Planck} channels 30, 100, 217 and 353 GHz by applying the EPSS--2D algorithm. Notice
that each plotted $C_{\ell}$ has been calculated from only one all-sky map of EPS, following the method
explained in \S 2 and with the assumptions made in \S 3. However, for giving a confidence interval, we also plot
the $\pm 1\sigma$ levels around the mean, having performed 100 simulations of flat sky patches. As in
\citet{TE96}, TO98 we represent the quantity $\delta T_{\ell}(\nu )=\sqrt {\ell (\ell +1)C_{\ell}/2\pi}$ (in
units of K).

\begin{figure}
\plotone{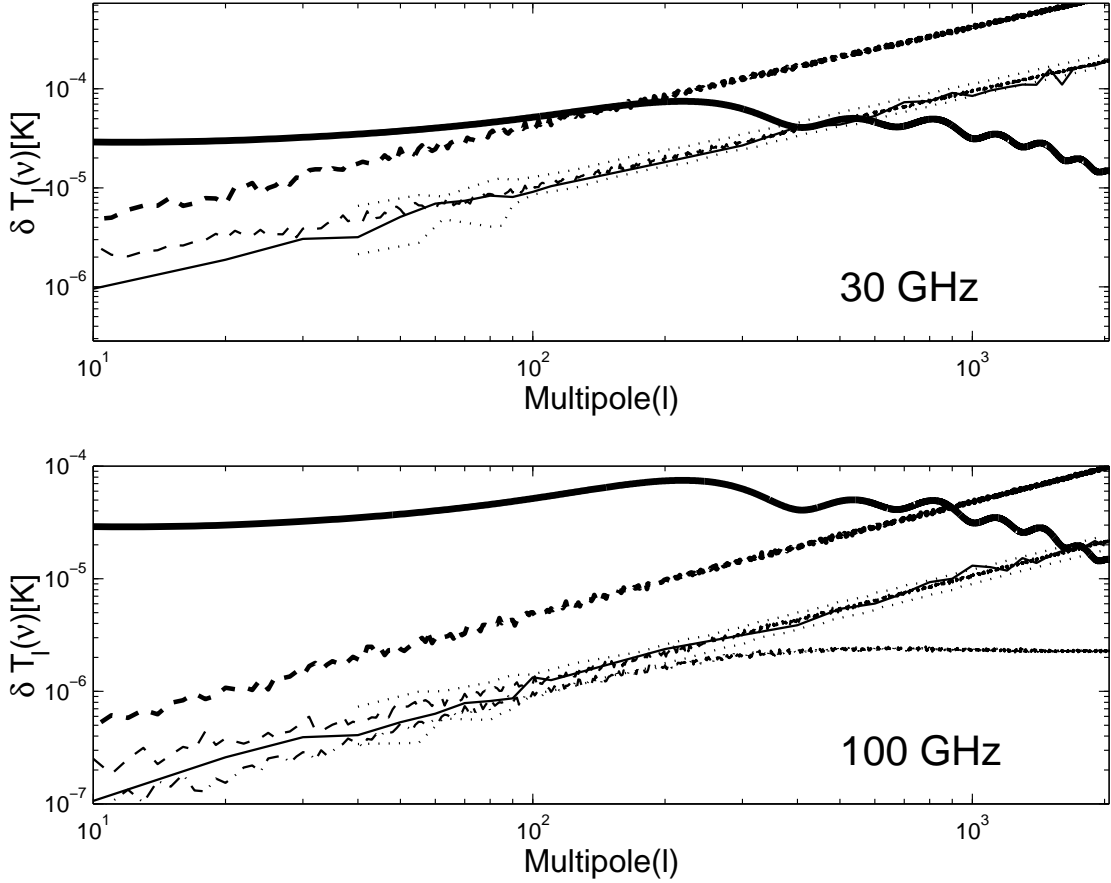} \caption{{\small Top panel: angular power spectrum
($\delta T_{\ell}= \sqrt{\ell (\ell + 1)C_{\ell}/2\pi}$) of
clustered and Poisson distributed EPS at 30 GHz. The thick dashed
line shows the $\delta T_{\ell}$ values recovered from the
simulated map without any source subtraction. The thin dashed line
represents the same quantity calculated after having subtracted
from the map all sources with $S\geq 1$ Jy. The thin continuous
line shows, for comparison, the prediction of TO98 obtained
applying the same detection limit for sources, $S_{lim}= 1$ Jy, as
before. The $1\sigma$ confidence level obtained by 100 simulations
on 2D sky patches is also shown (dotted lines). In this latter
case the largest angular scales (lowest multipoles) are not
sampled. Bottom panel: angular power spectrum of EPS at 100 GHz.
The meaning of the lines is the same as in the top panel. The
dot-dashed line in the lower right side of the panel represents
the predicted angular power spectrum of clustered high redshift
spheroids, calculated by the $w(\theta )$ of \citet{neg04a}. In
each panel, the thick continuous line represents, for comparison,
the primordial CMB power spectrum calculated for a flat
$\Lambda$CDM model.}}
\end{figure}

\begin{figure}
\plotone{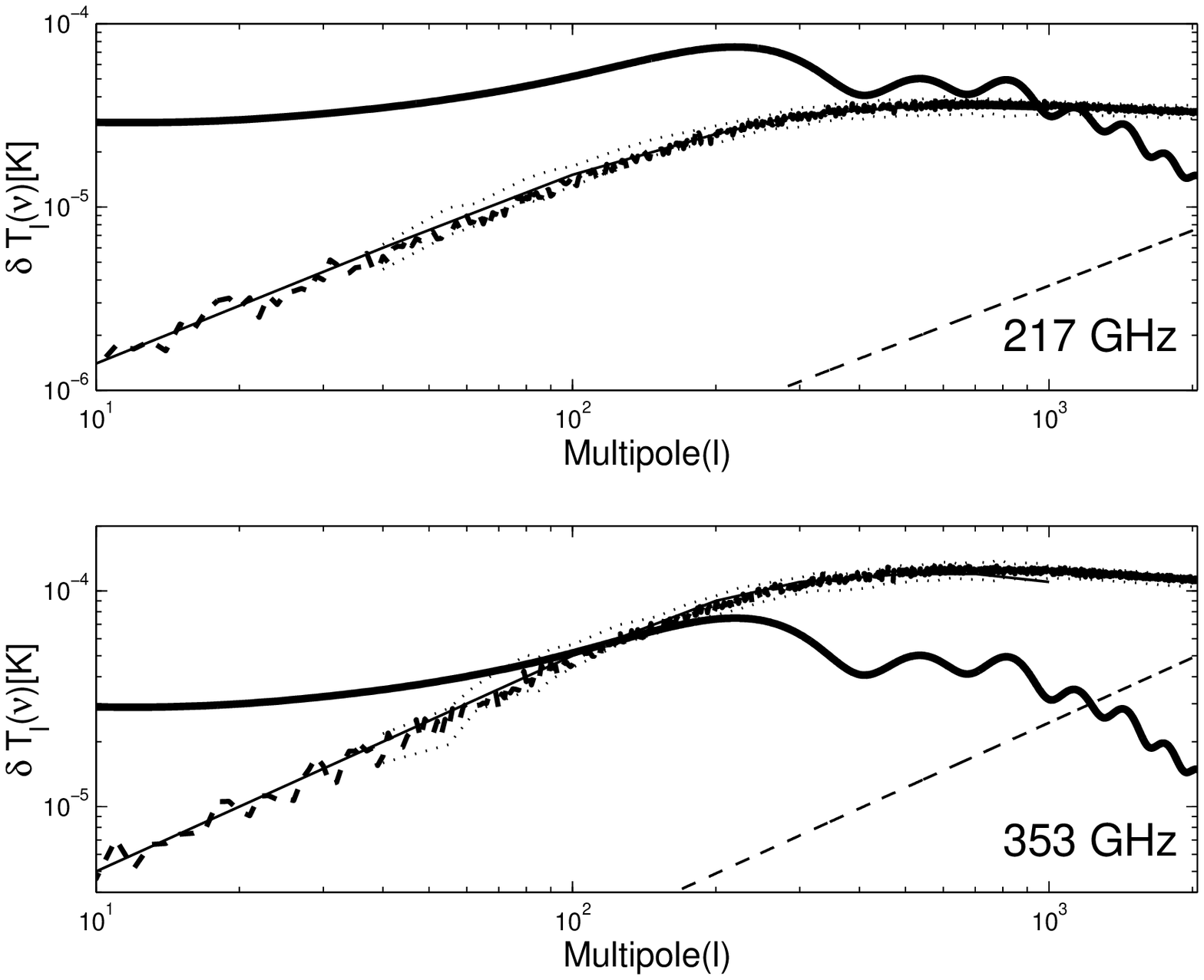} \caption{{\small Top panel: angular power spectrum
($\delta T_{\ell}= \sqrt{\ell (\ell + 1)C_{\ell}/2\pi}$) of
clustered and Poisson distributed EPS at 217 GHz. The thick dashed
line shows the $\delta T_{\ell}$ values recovered from the map by
applying the same detection limit, $S_{lim}$, for EPS as in the
theoretical predictions obtained by \citet{per03} for the case
$M_{halo}/M_{sph}=100$ (thin continuous line). The $1\sigma$
confidence level obtained by 100 simulations on 2D sky patches is
also shown (dotted lines). The thin dashed line shows the $\delta
T_{\ell}$ values obtained with Poisson distributed sources with
the same detection limit as in \citet{per03}. Bottom panel:
angular power spectrum of clustered EPS at 353 GHz. The meaning of
the lines is the same as in the top panel. In each panel, the
thick continuous line represents, for comparison, the primordial
CMB power spectrum calculated for a flat $\Lambda$CDM model.}}
\end{figure}

As discussed in \S 3.2.1, due to the relatively large beam area (i.e., bright detection limits) of current as
well as future all-sky CMB anisotropy experiments, we can be confident in simulating 2D all--sky maps by using
only ``flat''--spectrum cm selected radio EPS at $\nu\leq 100$ GHz. The angular power spectrum, $P(k)_{cl}$,
determined by using the angular correlation function of \citet{lwl97} has been used, as before, given that the
parent population of bright sources, i.e. blazars, is the same one as at 5 GHz. We rely on this choice since
current earth--borne as well as satellite surveys are confirming that other source populations, i.e. inverted
spectrum sources, are almost not sampled at bright fluxes if $\nu\leq 100$ GHz \citep{ben03b}.

From Figure 4 it is clear that the recovered $C_{\ell}$s of clustered EPS are still compatible with the ones
recovered from a Poisson distribution of sources \citep{TO98}. The contribution of clustered EPS sources to
temperature CMB fluctuations at these frequencies is {\it found to be negligible at all angular scales}, ``if
sources are not subtracted down to fluxes well below the detection limit of the survey, thus greatly decresing
the Poisson fluctuations, while the contribution arising from clustering is only weakly affected'' \citep{TO98}.
In fact, if only bright sources at $S> 1$ Jy are subtracted out from the map -- e.g., the detection limit of the
WMAP survey -- the $C_{\ell}$ of clustered EPS still match very well the predictions of TO98. These results are,
again, in agreement with the well known observational result: ``the wide redshift range of radio sources washes
out much of the clustering signal'' (BW02). From the same Figure 4 it is possible to appreciate a small excess
at the largest angular scales in the estimated $C_{\ell}$s, of clustered EPS, if compared to the ones estimated
by TO98. This is in agreement with the well known outcome that the ratio of clustering--to--Poisson fluctuations
increases with increasing angular scale, i.e. at decreasing multipole number \citep[see, e.g.,][]{dez96,neg04a}.

The $C_{\ell}$ of clustered dusty proto-spheroidal galaxies at 100 GHz (see Figure 4) confirms previous findings
of \citet{neg04a} on the confusion noise: the Poisson term due to radio selected sources is always the dominant
one at all multipoles at $\nu\leq 100$ GHz. On the other hand, at $\nu \geq 150$ GHz spheroids probably dominate
the fluctuations due to EPS.\footnote{We have checked that our current predictions on confusion noise and the
ones by \citet{neg04a} are in very good agreement at each frequency: the differences are always $\leq 5$\%. As
an example, at 100 and 143 GHz we find $\sigma_{C,sph}=2.5,\, 5.6$ mJy, respectively, at the Planck resolution,
to be compared with the values $\sigma_{C,sph}=2.6,\, 5.7$ mJy quoted by \citet{neg04a} and by using the same
detection thresholds. The contribution of spheroidal galaxies is almost insensitive to the adopted flux limit,
because of the very steep counts at bright fluxes.}

In Figure 5 we plot our current outcomes at 217 and 353 GHz by using the EPS counts of spheroidal galaxies of
\citet{gra01} (see \S 3.2.2) and by applying the $w(\theta )$ discussed in \citet{per03}. In this case, due to
the very strong angular correlation function detected by SCUBA \citep{pea00}, for recovering the input $w(\theta
)$, we have to force the brightest fluxes to fall in the highest density pixels. From Figure 5 it is clear that
a very good agreement is found between the recovered $C_{\ell}$s and the theoretical predictions of
\citet{per03}, at all multipoles. The small scatter around the average value, displayed by the $\pm 1\sigma$
confidence intervals, shows that we can be confident in the $C_{\ell}$s recovered from each simulated map. As
previously noticed, at 217 GHz (3rd {\it Planck} HFI channel) a comparable contribution to bright EPS counts is
expected either from spheroidal galaxies and from radio selected ``flat''--spectrum sources. Anyway, given that
this latter population is much less clustered and, moreover, thanks to the dilution of the clustering signal --
due to their flat luminosity function and broad redshift distribution \citep{dun90,b&w02} -- the contribution of
``flat''--spectrum EPS falls below the one coming from spheroids \citep[see, e.g.,][]{neg04a}. At 353 GHz,
spheroids starts to dominate the counts at very bright fluxes and, thus, if they are strongly clustered as
suggested by SCUBA surveys \citet{pea00} their $C_{\ell}$s are no doubt the dominant ones. As for other far--IR
selected source populations, they give a small to negligible contribution to CMB temperature fluctuations, since
they are much less clustered (see \S 3.2.2; \citet{tof04}). Therefore, we can be sufficiently confident in our
current results.


\subsection{The excess power detected at arcmin scales, i.e. $\ell\geq 2000$}

Recent CMB experiments, aimed at probing arcmin angular scales, i.e. DASI, BIMA amd CBI, respectivey
\citep{hal02,daw02,mas03,read04}, have tentatively detected a signal at high multipoles well in excess on
predictions for primordial CMB temperature anisotropies. The interpretation of these outcomes is still open.
Possible interpretations are in terms of the thermal Sunyaev--Zel'dovich (SZ) \citep{SZ80} signal associated to
the formation of spheroidal galaxies \citep{dez04} or in terms of the SZ effect in galaxy clusters
\citep{read04}. However, all these interpretations require very high values of the $\sigma_8$ normalization. On
the other hand, the emission from advection-dominated accretion flows (ADAF) around supermassive black-holes in
early-type galaxies is found to be not sufficient to explain the detected excess \citep{pier04}.

Undetected EPS can represent an alternative (or, at least, a partial) explanation to the excess.
Given that extragalactic sources are, very probably, the dominant contributor to CMB
fluctuactions at these small angular scales, they have to be carefully
subtracted out. And, indeed, all the quoted works devoted a great effort to this subject.
In Figure 6 (bottom panel) we plot, as an example, the $C_{\ell}$ due to Poisson
distributed EPS, as estimated by \citet{hal02} and our prediction at 30 GHz by using the
TO98 model counts and a detection limit $S\simeq 50$ mJy for sources. The very good agreement
between our current results and those estimates confirms, again, that the TO98 model still
holds well, down to fluxes of tens of mJy at $\nu\leq 30$ GHz (see Section 3.1). Moreover,
this agreement also implies that source clustering {\it cannot} add a relevant contribution to the observed signal, as previously discussed.

\begin{figure}
\epsscale{.70}\plotone{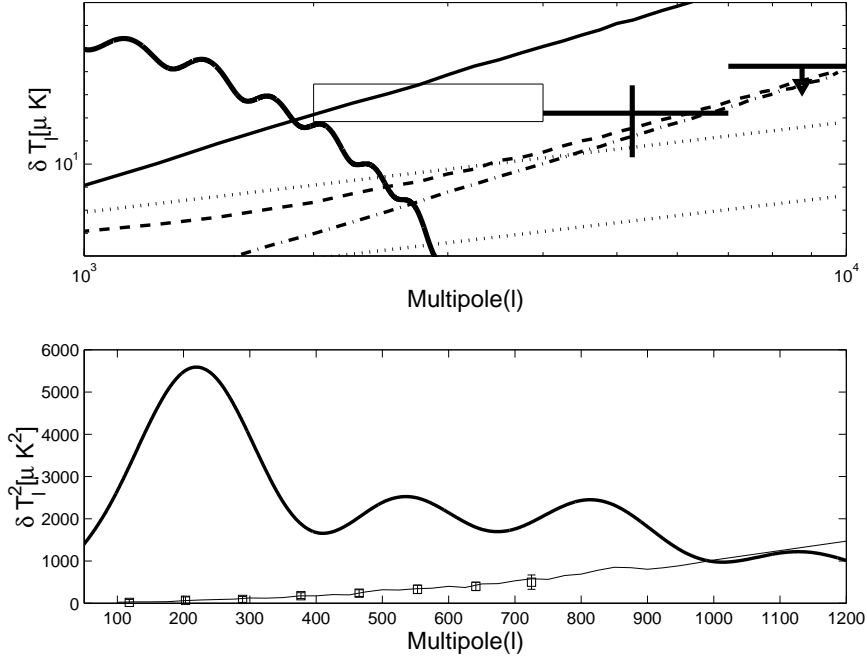} \caption{{\small Top panel:
contributions of EPS to the excess power detected at arcmin
angular scales by the CBI (box) and BIMA (cross and upper limit)
experiments. The thickest curve on the left side of the panel
represents the primordial CMB power spectrum. The thick continuous
line -- to be compared with the CBI excess -- show our estimate of
the total power, i.e. Poisson+clustering term, due to undetected
(residual) EPS at 31 GHz by adopting the $S_{lim}=$3.4 mJy of
\citet{read04}. The two dotted lines -- \citet{lwl97} $P(k)$:
upper one; \citet{b&w02} $P(k)$: lower one -- refer to the extra
power given by the pure clustering term with $S_{lim}=3.4$ mJy.
The thick dashed and dot-dashed lines -- to be compared with the
BIMA data -- refer to the total power (thick dashed) and to the
pure Poisson term (dot dashed) due to undetected radio sources and
to a detection limit of $\simeq 400$ $\mu$Jy. Bottom panel: DASI
data on the angular power spectrum of undetected sources (filled
squares). The continuous thin line show our current estimate by
using the same source detection limit as in \citet{hal02}. The
thick curve represents the primordial CMB power spectrum.}}
\end{figure}

As discussed by \citet{dez04,tof04}, a quite substantial residual contribution of radio selected sources to the
CBI signal is difficult to rule out, whereas the residual contamination of the BIMA data should be, very likely,
smaller. All published estimates of EPS contamination of CMB maps at high multipoles have used a Poisson
distribution of extragalactic sources. By our simulation code and the TO98 model counts we can now assess, with
good confidence, the excess contribution due to clustered EPS sources. In Figure 6 we plot the total noise
(Poisson+clustering; continuous line) and the pure clustering term (dotted lines) given by radio selected
sources following the TO98 source counts. We model the clustering term by using the $w(\theta )$ of
\citet{lwl97} and BW02 as discussed in Sections 2 and 3.


The top panel of Figure 6 shows that undetected sources below $S_{lim}=3.4$ mJy, the most recent detection limit
quoted by \citet{read04}, can easily explain the excess. If we guess that radio selected sources do cluster down
to very faint flux limits ($S_{min}=0.01$ mJy) following the $w(\theta )$ of \citet{lwl97}, then the clustering
term should give a non negligible contribution ($\Delta T\simeq 9$-$10\,\mu$K) to the total fluctuations. On the
other hand, if we adopt the $w(\theta )$ of BW02, which can represent a more realistic approximation to the
clustering properties of faint undetected sources in the CBI fields, since we are dealing with very faint
sources not sampled by the NRAO-VLA survey, the additional power due to clustering ($\Delta T\simeq
3$-$4\,\mu$K) is indeed very small in comparison with the Poisson term.


As for the BIMA data plotted in Figure 6, our current estimates do indicate that Poisson distributed and
clustered EPS selected at radio frequencies are the dominant contributor to the detected excess, by adopting the
effective detection limit $\simeq 400\,\mu$Jy, below which the counts of detected sources sink down
\citep{tof04}, and by keeping the same assumptions as before for their clustering properties. On the other hand,
at the very faint flux limits (down to $S\simeq 0.1$ mJy) probed by the BIMA experiment other source populations
-- undetectable at brighter fluxes -- show up. The detailed analysis of \citet{tof04} shows that high-redshift
dusty proto-spheroidal galaxies in the main phase of star formation \citep{gra04} are contributing to arcmin CMB
fluctuations even at frequencies as low as $\sim 30$ GHz. Moreover, if these sources show the mm excess detected
in several Galactic clouds as well as in nearby galaxies, their contribution to CMB fluctuations would be higher
and, when summed in quadrature to the contribution of radio selected sources, would imply a signal only
marginally consistent with the BIMA data.

\subsection{A realistic all--sky map of EPS at 70 GHz}

As an application, we have combined current WMAP data on EPS \citep{ben03b} with our 2D simulation code for
creating a {\it realistic map} of EPS in the submillimeter domain. Obviously, it is necessary to follow a
four--step process: first of all, we use the EPSS-2D code for simulating an all--sky map at 70 GHz by using the
same assumptions made in \S 3.2.1 and distributing only sources with fluxes $S\leq 1$ Jy. This flux density
roughly corresponds to the source detection limit of the WMAP K and Q channels, in which the WMAP survey reaches
the highest number of detections. Subsequently, the flux densities, $S_i$, of all WMAP sources have been
converted to the corresponding values at {\it Planck} LFI frequencies, by using the spectral indexes published
by \citet{ben03b}. Among all fluxes, the ones measured at 61 GHz -- by the WMAP V channel -- have been used as
the reference values for converting them to 70 GHz, given that this is the WMAP channel closest to the frequency
of the map. In the case of a no detection in the V channel, the flux measured in the Q channel has been used;
alternatively, the density flux of the K channel has been exploited, in such a way that we could convert all
fluxes to our frequency.

Then, every source in the WMAP 1--year catalogue has been placed in that specific pixel of the HEALPIX
(nside=2048) pixelization scheme corresponding to the published Galactic coordinates of the source. Notice that,
after having determined the position of each WMAP source, i.e. the corresponding HEALPIX pixel, we have {\it put
to zero} the flux previously simulated by our code in that particular pixel. Finally, we assign to each
``zero--flux'' pixel the flux (or, equivalently, the temperature) of the WMAP source corresponding to that
position. Moreover, we have also checked that all the simulated fluxes which fall inside a ring -- of the
dimension of the WMAP FWHM -- around each WMAP source are not greater than the corresponding error in flux of
the WMAP 1-year EPS catalogue. This choice guarantees that we can exactly recover the fluxes of all the
brightest -- WMAP detected -- sources from our map, shown in Figure 7. This map is a very realistic realization
of the EPS sky at 70 GHz given that it combines the WMAP catalogue (bright detected sources) and realistic 2D
simulations (faint sources), as discussed before. Moreover, as soon as the 2-year WMAP catalogue shall be
available, this same map, as well as other maps of EPS at other frequencies $\nu\leq 100-150$ GHz, can be easily
and rapidly updated.

\begin{figure}
\epsscale{.60} \plotone{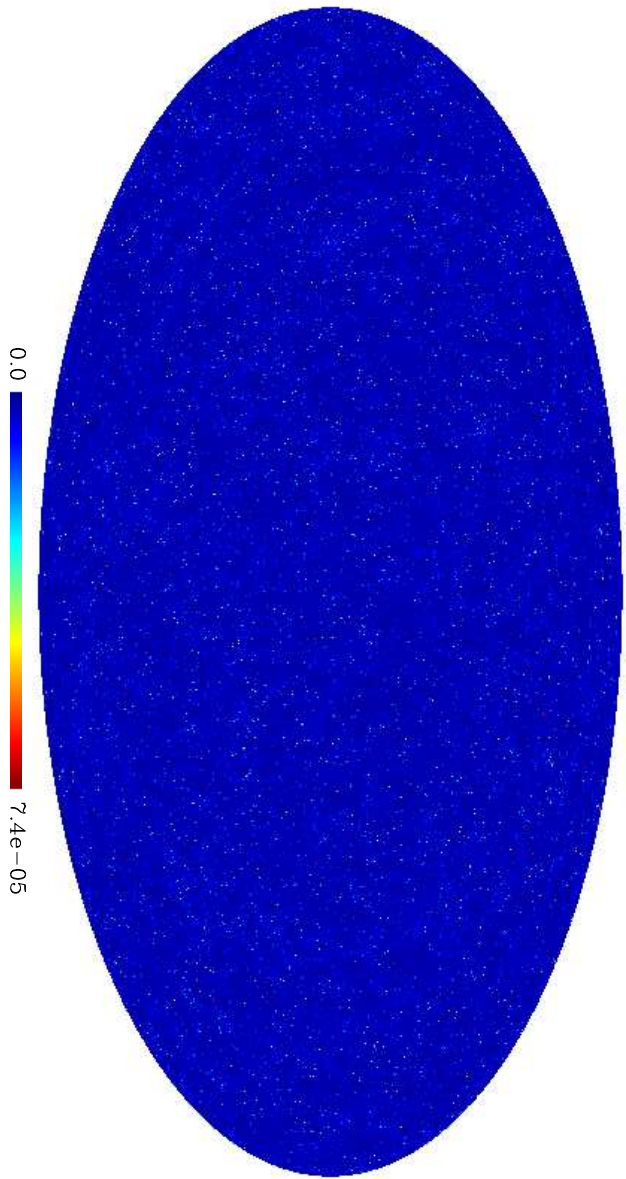} \caption{{\small Full sky map of
EPS at 70 GHz represented in the galactic coordinate system. In
the chosen illustration fluxes have been converted to
thermodynamic temperature fluctuations, $\Delta T/T$, where
$T\equiv T_0=2.725 K$ \citep{mat99}. The maximum value in the map,
corresponding to the brightest source, n.181 in the 1--year WMAP
Catalogue of \citet{ben03b}, is $\Delta T/T=0.2745$. The map has
been smoothed by a 13.5$^\prime$ FWHM beam, the angular resolution
of the {\it Planck} 70 GHz channel. In this representation, we
selected a scale with a maximum value $\Delta T/T_{max}=\langle
\Delta T/T\rangle +10\langle (\Delta T/T)^2\rangle^{1/2}$, to
better show the fluctuation field due to sources.}}

\end{figure}

\section{CONCLUSIONS}

We have presented predictions on the angular power spectrum of EPS by 2D simulated maps in which extragalactic
sources are distributed with correlated positions in the sky. The code, named EPSS-2D, is able to simulate both
sky patches and all--sky maps taking short CPU process times in a common Workstation (see \S2.2.1). We want to
stress again that we adopted a purely {\it phenomenological} approach given that our main purpose was the
definition of a fast and flexible tool for simulating 2D maps of EPS, under the most general assumptions on
source counts and on the angular correlation functions of point sources. The present outcomes are, obviously,
{\it model dependent}. On the other hand, they try to take into account all current data on EPS counts as well
as on source clustering. All these simulated maps can be very useful for studying the effect of clustered EPS,
in view of estimating the CMB angular power spectrum and bispectrum.

The main results of this paper may be summarized as follows:

1) 2D maps are created by first making a Poisson density field, corresponding to the differential counts of EPS,
$N(S)$, which have been determined or estimated at the frequency under study; subsequently, we modify -- in the
Fourier space -- this ``white noise'' density field according to some angular power spectrum, $P(k)$, the most
reliable one for that specific source population, whose differential counts have been used; as a final step, we
distribute fluxes on the density field map, under the condition that $n$ fluxes -- taken from the differential
counts -- fall in a pixel whose number of sources is $n$. This method, discussed in detail in \S 2, proves
``safe'' given that we are always able to recover the input source counts and the input $P(k)$. Moreover, the
proposed method allows us to estimate also the contribution of clustered EPS to the CMB bispectrum
\citep{arg03}.

2) The method allows, in principle, the simulation of as many EPS populations as needed, provided that their
contribution to the differential counts at a given frequency is known. On the other hand, in almost all the
cases it will be sufficient to only distribute in the sky those EPS coming from the dominant source population
at a given frequency, thus making the simulation simpler and faster. This is generally the case, except in some
very particular frequency range, in which number counts -- at fluxes relevant for a particular experiment -- are
contributed by a comparable number of sources coming from two (or more) source populations which show different
clustering properties (see \S 2.2.2).

3) By using the Toffolatti et al. cosmological evolution model for EPS and by applying the $w(\theta)$ of
\citet{lwl97} we have estimated the temperature angular power spectrum of clustered EPS at WMAP and {\it Planck}
LFI frequencies. Our current results confirm that the extra power due to clustering of EPS is {\it always small}
in comparison with the Poisson term in all cases for which no subtraction of bright sources is applied (see \S
4.1). As extensively discussed in the body of the paper, the present outcome is in full agreement with all the
main studies on the subject.

4) On the other hand, we find that at frequencies $\nu\geq 150-200$ GHz, the contribution of the clustering
term, $\sigma_C$, to the total confusion noise can be, very probably, the dominant one, thus confirming previous
theoretical predictions. This result is mainly determined by the very steep slope of EPS counts at sub--mm
wavelengths combined with the strong clustering signal inferred for high redshift spheroidal galaxies and SCUBA
sources \citep{pea00,per03,neg04a}.

5) Our current simulations allows us to also estimate the contribution of clustered EPS to the excess signal
recently detected at arcmin scales by the CBI and BIMA fields. Our current findings, which -- as stated before
-- are model dependent, suggest that clustering of EPS can give a small but not negligible contribution to this
excess.

6) We present here, as an example, a realistic all--sky map of clustered EPS sources at 70 GHz. This map has been
obtained by combining the WMAP point source catalogue, converted to 70 GHz, and a map of fainter EPS at the same
frequency, simulated by the EPSS--2D code. As previously discussed, this map and other ones at WMAP and {\it
Planck} LFI frequencies are easily simulated, thanks to the currently available WMAP 1--year data. The lack of
data at higher CMB frequencies makes a map of this kind much more uncertain.

Finally, we remind that we can very easily update our current analysis and results: e.g., by taking into account
more source populations, for providing more accurate estimates on the $C_{\ell}$s of EPS. As for the simulated
all-sky maps, we can easily update them by the EPSS--2D code and by exploiting the 2-year WMAP source catalogue
in the next future.

It is important to stress again that current and future applications of these maps are manifold: first of all,
they allow the direct estimate of $C_{\ell}$s due to EPS, under very general assumptions; they shall be useful
in testing the capability of new algorithms for the detection of EPS in presence of correlated positions in the
sky; they shall also allow us to test the efficiency of component separation techniques; eventually, they can be
of some help in the study of non Gaussian signatures in residue CMB maps. Therefore, we intend to make publicly
available the maps created by our EPSS--2D simulation code. We hope that they shall be useful to all the
research community devoted to the development of current as well future CMB anisotropy experiments.



\acknowledgments

We thank an anonymous referee for useful comments and suggestions which helped us in improving the final
presentation of this paper. We are grateful to J.M. Diego, E. Mart\'\i{nez}--Gonz\'alez, P. Vielva and R.B.
Barreiro for very stimulating discussions as well as for some useful remark on an early version of this paper.
We are indebted to C. Burigana for a useful suggestion which helped in the final definition of the all--sky map
at 70 GHz as well as for the .ps image of the same map. We also thank: G.L. Granato for providing us with his
model counts of EPS at Planck HFI frequencies; M. Negrello for his useful collaboration in checking the
agreement between his theoretical predictions and our simulations on the angular power spectrum of EPS; G. De
Zotti for reading carefully the final version of the paper. We acknowledge partial financial support from the
Spanish MCYT under project ESP2002--04141--C03--01. JGN acknowledges a FPU fellowship of the Spanish Ministry of
Education (MEC). This work has used the software package HEALPIX (``Hierarchical, Equal Area and Isolatitude
Pixelization of the Sphere'', http://www.eso.org/science/healpix), developed by K.M. Gorski, E.F. Hivon, B.D.
Wandelt, J. Banday, F.K. Hansen and M. Bartelmann.

\end{document}